\documentclass{epl}

\newcommand{\pa}{\partial}
\newcommand{\la}{\lambda}
\newcommand{\de}{\delta}
\newcommand{\al}{\alpha}

\newcommand{\ga}{\gamma}

\newcommand{\myref}[1]{(\ref{#1})}
\usepackage{amssymb}
\usepackage{amsmath}
\usepackage{latexsym}

\title{Persistence distributions for non gaussian markovian processes}
\author{Jean Farago\inst{1}}
\institute{
  \inst{1} Laboratoire de Physique UMR CNRS 5672 - ENS de Lyon, 69364 Lyon cedex 07 France\\
}

\begin{document}

\maketitle

\begin{abstract}
We propose a systematic method to derive the asymptotic
behaviour of the persistence distribution, for a large class of 
stochastic processes described by a general Fokker-Planck equation in one
dimension. Theoretical predictions are compared to simple solvable systems
and to numerical calculations. The very good agreement attests the validity
of this approach.
\end{abstract}

\section{Introduction}

The concept of persistence has recently motivated a lot of
works, both experimentally and theoretically (see \cite{majumdar} and
references therein) : it appeared that this simple
concept --i.e. the probability $G(t)$ that a random variable $X(t)$ never goes
above a certain value (usually its mean value) during the whole time interval $[0,t]$-- hides a real
complexity, since its definition involves a complete
knowledge of the process over a large time interval. Consequently, the
persistence distribution gives information on the details of the
process, which are not redundant with the traditional statistical tools,
as for instance the correlation functions. Until now, the works was focused,
besides  discrete systems as spin systems \cite{godrechedornic}, 
essentially on gaussian processes, and the efforts concentrated on the influence
of the non markovian character of the process on the persistence
exponent. 
However, it was shown recently \cite{deloubrierehilhorst} that the concept of
persistence can also be physically relevant for non gaussian situations\footnote{There
  is a slight ambiguity associated with the term ``non gaussian'' ; in
  \cite{deloubrierehilhorst}, the process is non gaussian due to the non
  gaussianity of the noise ; but the process can be non gaussian, despite
  a gaussian noise, if an external force is present and not linear.}. Another physical situation  leads naturally to study
persistence in a non gaussian context : in nonlinear physics, the
combination of 
discreteness and nonlinearity gives rise to localisation phenomena (the
``breather-modes'' \cite{mckayaubry,flachwillis}), which are often  extremely pinned  where they appear; a
powerful tool to characterise the lifetime of these objects is provided by
the persistence behaviour of the energy density on a site. In weakly coupled
systems, the evolution of the energy density can  moreover be described by a non gaussian
markovian process, with energy dependent drift and diffusion coefficients.

In this letter, we introduce a general method to compute the persistence
behaviour for a large category of processes described by a one-dimensional
Fokker-Planck equation. Essentially, the method deals with systems where the
persistence  decreases  slower than an exponential law. We check the
validity of our predictions by a comparison with exact results and
numerical simulations.

\section{The method}
We consider $X(t)$ a continuous markovian stochastic process. By
definition, the
probability density function (pdf) $p(x,t)=Prob(X(t)=x)$ obeys the
Fokker-Planck equation
\begin{equation}
  \label{FP}
  \pa_tp=-\pa_x(A(x)p)+\frac{1}{2}\pa^2_{xx}(B(x)p)
\end{equation}
where the functions $A$ and $B$ characterise the dynamical process. Once
given a real value
$x_0\in\mathbb{R}$, the probability of (positive) persistence $G_+(x,t)$,
i.e. the probability that a particle, which has initiated its trajectory at
$x>x_0$ at time $t=0$, has never crossed $x_0$ up to the time $t$, obeys
the backward Fokker-Planck equation \cite{gardiner}
\begin{equation}
  \label{BFP}
  \pa_tG_+=A(x)\pa_xG_++\frac{B(x)}{2}\pa^2_{xx}G_+
\end{equation}
with the boundary conditions $G_+(x>x_0, t=0)=1, G_+(x=x_0,t>0)=0$.

For the sake of clarity, we will first temporarily restrict the discussion to 
cases where $B(x)=2$, and see later how to take into consideration
the general case. In addition, we assume that the stationary solution of \myref{FP},
$p_{st}(x)\propto \exp\int^xA(x')dx'$, is bounded but not necessarily
normalisable (it means just that the potential of the force doesn't go to
$-\infty$).

The  Fokker-Planck equation \myref{BFP} with constant diffusion
coefficient can be mapped on a Schr\"odinger equation \cite{risken}, by
defining $\psi_+(x,t)=G_+(x,t)\sqrt{p_{st}(x)}$ : it reads
\begin{equation}
  \label{schro}
  \pa_t\psi_+=\pa_{xx}^2\psi_+-(\frac{A'}{2}+\frac{A^2}{4})\psi_+.
\end{equation}
The effective potential $V(x)$ of this
equation is defined on $[x_0,+\infty[$ by $V(x>x_0)=A'(x)/2+A^2(x)/4$
\textit{and} $V(x_0)=\infty$; the last precision  ensures the boundary
conditions $\psi_+(x_0,t)=0,\ \forall t$ \cite{cct}. It can be shown that
the spectrum of this Schr\"odinger operator $[-\pa_{xx}^2+V(x)]$ is only located in $]0,+\infty[$
 (due to the boundedness of $p_{st}$), and we know from quantum
mechanics that it  usually consists of  a discrete part
$\{\lambda_1<\la_2<\ldots<\la_p\}$ (with normalisable eigenfunctions $\psi_{\la_j}$), surmounted by a continuous part
$]\la_c,+\infty[$ (with non normalisable eigenfunctions $\psi_\la$). The
value of $\la_c$ is usually easy to determine (since $\la_c$ is more or less related to
$\lim\inf_{+\infty} A^2(x)/4$).

\bigskip

We obtain therefore the persistence distribution $G_+$ as
\begin{align}
  G_+(x,t)&=\frac{1}{\sqrt{p_{st}(x)}}\left\{\sum_{i=1}^pe^{-\la_it}\psi_{\la_i}(x)B_{\la_i}+\int_{\la_c}^\infty d\la \ e^{-\la t}\psi_\la(x)B_\la\right\}
\end{align}
where $B_\la=\int_{x_0}^\infty
dx\ \psi_\la(x)\sqrt{p_{st}(x)}$

The asymptotic temporal dependence of $G_+$ can therefore exhibit three different
scenarios: first, if  a discrete family of eigenvalues exists
for the Schr\"odinger operator,  the tail of
distribution is proportional to $\exp(-\la_1 t)$, and the appropriate
quantity which describes the asymptotic behaviour is the persistence time
$1/\la_1$. Consequently, the possibility of calculating this time is simply
related to the ability of estimating the lowest eigenvalue of the corresponding
Schr\"odinger problem.

If we consider now the cases without bound states, there are two
possibilities : if $\la_c$ is different from zero (existence of a gap), 
 the behaviour of $G_+$ will not be simple
\textit{a priori} :  $G_+(t)\propto f(t)\exp(-\la_c t)$ (for
large $t$), where $f(t)$ is unknown. An example of such a situation is the
case of a particle experiencing a constant force toward $0$ ($A<0$ constant). In that simple
case, we have $f(t)\propto t^{-3/2}$.  Our method does not deal with these
cases, which we could call marginal because the force acting on the
particle neither grows nor disappears at infinity.

Finally, if $\la$ is equal to zero (``gapless'' situation), we will show
that there is  a general
procedure for obtaining the asymptotic behaviour of the persistence, whatever
 the $A(x)$ under consideration. First, the formal formula
\begin{align}
  G_+(x,t)&=\frac{1}{\sqrt{p_{st}(x)}}\int_{0}^\infty d\la \ e^{-\la t}\psi_\la(x)B_\la
\end{align}
shows that we have to determine the $\la\rightarrow 0$ behaviours of $B_\la$
and $\psi_\la$ in order to get the $t\rightarrow\infty$ limit. The
cornerstone of our method is that it is possible to determine this
behaviour, thanks to a peculiarity of the Schr\"odinger problem under
consideration:   the ``$\la=0$
eigenfunction'' is known, because  it is related to the stationary solution of the
corresponding  Fokker-Planck equation. The quotation are required, because
this eigenfunction is highly diverging at $t\rightarrow\infty$ : strictly
speaking, this $\psi_{\la=0}$ does not belong to the spectrum. Before going
further, let us show how $\psi_0$ is constructed. One of the functions
fulfilling $\psi''-(A'/2+A^2/4)\psi=0$  is just
$\sqrt{p_{st}}$. Another (not proportional) is obtained by a simple
quadrature, $\sqrt{p_{st}}\int^x\ dx'p_{st}^{-1}(x')$,
and clearly diverges at infinity. As the ``$\la=0$'' eigenfunction must
fulfil the boundary conditions at $x_0$, we can write it as
\begin{align}
  \psi_0(x)=\sqrt{p_{st}(x)}\int_{x_0}^x\ \frac{dx'}{p_{st}(x')}
\end{align}
This ``eigenfunction'' is  useful, since it is reasonable to assume that
despite the diverging character of $\psi_0$, there is a continuity property of
$\psi_\la(x)$ with respect to $\la$, which leads to  the limit $\forall x,
\lim_{\la\rightarrow 0}\psi_\la(x)=\psi_0(x)$. However, there will of
course   not be
any related property of uniform convergence.

 With this assumption, which could
presumably be proved using rigorous mathematical arguments, we  have the typical
portrait of $\psi_\la(x)$ for vanishing $\la$, over the whole range
$[x_0,\infty[$ : ``far'' from the potential region, 
$\psi_\la(t)$ must become plane waves : $\psi_\la(x)\rightarrow
W_\la\cos(\sqrt{\la}x+\phi_\la)$. The phase factor $\phi_\la$ is assumed to
have a limit $\phi_0$ when $\la$ goes to zero, but its precise knowledge is
useless for the persistence. On the contrary,  it is
important to know the $\la$-dependence of the coefficient of
proportionality $W_\la$. The requirement that
$\psi_\la(x)$ must be normalised in the sense $\int_{x_0}^\infty
dx\psi_\la(x)\psi_{\la'}(x)=\de(\la-\la')$ imposes that $W_\la=C^{st}\times
\la^{-1/4}$ \cite{landau}.

On the other hand, in the region where the potential is substantially
different from zero, the differential equation for $\psi_\la$ is well approximated by the
equation $\psi''-V\psi=0$ ; so in this region we can expect therefore 
\begin{align}
  \psi_\la(x)\sim A_\la\psi_0(x)
\end{align}
where, once again, the $A_\la$ is important to determine, by estimating the location of the crossover of the two limiting
behaviours. This location is naturally
characterised by the abscissa $x_\la$ given by the balance 
\begin{align}\label{xlbda}
  V(x_\la)\sim\la
\end{align}
Nevertheless  this condition will
only hold if the computed $x_\la$ is larger than a wavelength
$\sim1/\sqrt{\la}$ corresponding to the minimum typical length the eigenfunction needs to
join the maximum of his asymptotic behaviour. In other words, if \myref{xlbda} gives a
$x_\la$ which diverges more slowly than $1/\sqrt{\la}$, one has to choose
$x_\la\sim1/\sqrt{\la}$ instead. 
 In some particular cases, the determination of $x_\la$ is easy to
 achieve, since $\la$ is
small. At this point $x_\la$, the continuity of $\psi_\la$ imposes that the
magnitude of the two branches must be the same : this condition,
\begin{align}\label{Albda}
  A_\la\psi_0(x_\la)\sim \la^{-1/4},
\end{align}
gives the $\la$-dependence of $A_\la$.

\bigskip

The aim is therefore achieved, because $B_\la$ can be evaluated with
the approximation
\begin{align}\label{Blbda}
  B_\la\approx
  A_\la\int_{x_0}^{x_\la}dx\sqrt{p_{st}(x)}\psi_0(x)+\la^{-1/4}\int_{x^\la}^\infty dx \sqrt{p_{st}(x)}\cos(\sqrt{\la}x+\phi_\la).
\end{align}
The relative importance of the two terms must be checked in each case. Moreover, further simplifications can be made ; for instance,
if $\int^\infty\sqrt{p_{st}}$ is converging, the second integral of
\myref{Blbda} is equivalent to
\begin{align}
  C^{st}\times\la^{-1/4}\int_{x^\la}^\infty dx \sqrt{p_{st}(x)}.
\end{align}
 The final calculation of the persistence behaviour is achieved by the
means of  saddle point expansion of the integral 
\begin{align}
  G_+(x,t)\propto \int_0^\infty d\la\  e^{-\la t}A_\la B_\la
\end{align}
 around $\la=0$.

\bigskip

Let us consider now the general case, i.e. $B(x)\neq 2$. The change of
variable defined by $dy/dx=\sqrt{2/B(x)}\equiv g$ leads to a new Fokker-Planck
equation for the distribution $p(y)=p(x)dy/dx$, with the new coefficients
$\widehat{B}(y)/2=1$ and $\widehat{A}(y)=A(y)g(y)+\pa_y g/g$. As a function
of $y$, we have then to compute the characteristic abscissa
$y_\la$. It is however more convenient to compute the
corresponding abscissa in the original coordinates, i.e. the $x_\la$
defined as the solution of
\begin{align}
\la &\sim
V(x_\la)=\frac{1}{2}\sqrt{\frac{B(x_\la)}{2}}\frac{d\widehat{A}}{dx}(x_\la)+\frac{1}{4}\widehat{A}^2(x_\la)\\
\mbox{where}\ \ \widehat{A}(x)&=\sqrt{\frac{2}{B(x)}}\times\left(A(x)-\frac{1}{4}\pa_xB(x)\right).
\end{align}
The $\psi_0$ eigenfunction can also be expressed in the original $x$
coordinate as
\begin{align}
  \psi_0(x)=\sqrt{p_{st}(x)\sqrt{B/2}}\int_{x_0}^x\frac{2dx'}{B(x')p_{st}(x')}.
\end{align}
 With this change of variables, the
coefficient $A_\la$ is given by $A_\la\psi_0(x_\la)\sim\la^{-1/4}$. The
same translation can be performed on $B_\la$, leading to
  \begin{multline}
    B_{\la}=A_\la\int_{x_0}^{x_\la}dx(2/B(x))^{1/4}\sqrt{p_{st}(x)}\psi_0(x)\\
+C^{st}\times\la^{-1/4}\int_{x_\la}^\infty dx (2/B(x))^{1/4}\sqrt{p_{st}(x)}\cos(\sqrt{\la}y(x)+\phi_0)
  \end{multline}
The presence of $y(x)$ in the cosine is not a problem in general, because only its
asymptotic behaviour is needed, which is often easily computed.

In summary, the cases where the diffusion coefficient $B(x)$ depends on $x$
are not a handicap to the procedure proposed in this letter.

\section{Examples}

To confirm the above procedure, we will first consider exactly solvable examples.
The simplest is probably the free brownian motion, i.e. $A=0,B/2=1$. In
that case, $V=0$ and $x_\la$ behaves as $1/\sqrt{\la}$ (the
prescription \myref{xlbda} giving $x_\la=0$, we have to choose the
``minimal divergence'' $1/\sqrt{\la}$) ; moreover, $\psi_0(x)\sim x$, and
$p_{st}=C^{st}$. Consequently, $A_\la\sim\la^{1/4}$, $B_\la\sim\la^{-3/4}$
and the persistence goes like $t^{-1/2}$, as the exact result states.

\medskip

To check the theory on a less trivial situation, let us consider the cases
$B/2=1,A=-\nu x^{-1}$ with $\nu>0$. The
Schr\"odinger potential is proportional to $x^{-2}$, and it is easy to
verify that, for large $x$,  $p_{st}\propto x^{-\nu}, \psi_0\propto x^{\nu/2+1},
x_\la\propto \la^{-1/2}, A_\la\propto \la^{(1+\nu)/4}, B_\la\propto
\la^{(\nu-3)/4}$. This gives the result $G_+(x,t)\propto t^{-(1+\nu)/2}$.
 The Schr\"odinger problem can actually be solved exactly,
 because the eigenfunctions $\psi_\la$ are proportional to
 $\sqrt{x}(J_{(\nu+1)/2}(\sqrt{\la}x)Y_{(\nu+1)/2}(\sqrt{\la}x_0)-J_{(\nu+1)/2}(\sqrt{\la}x_0)Y_{(\nu+1)/2}(\sqrt{\la}x))$ \cite{abramowitz}. The exact result matches our prediction (and the limit $\nu\rightarrow 0$ gives the free diffusion result).

\medskip

We turn now to situations for which  the knowledge of the
spectrum is unknown. For instance, consider the cases $A=-\nu x^{-\al},B/2=1$, with
$\nu>0,\al>0$. 

If we consider first the cases $\al>1$, we have then (for large $x$)
$p_{st}\propto\exp(\nu x^{1-\al}/(\al-1))$, $\psi_0\propto x$, $V\propto
x^{-\al -1}$, $x_\la\propto \la^{-1/(\al+1)}\leadsto x_\la\propto \la^{-1/2}$,
$A_\la\propto\la^{1/4}$, $B_\la\propto\la^{-3/4}$. It gives $G(x,t)\sim
t^{-1/2}$. It is interesting to note that the potential slope is inefficient to modify the
persistence behaviour of the particle, and that one obtains a discontinuity
of the exponent as a function of $\al$ (when $\al\rightarrow 1$).

 For cases $0<\al<1$, the
situation is completely different :  $\psi_0(x)\sim x^\al\exp(\nu
x^{1-\al}/2(1-\al))$, $V\sim x^{-2\al}$ and $x_\la\sim\la^{-1/2\al}$. It
leads to $A_\la\sim\la^{1/4}\exp(-(\la/\la_0)^{-(1-\al)/2\al})$, $B_\la\sim
A_\la \la^{-(1+\al^{-1})/2}$, where $\la_0\propto4(\nu/2(1-\al))^{(1-\al)/2\al}/\nu^2$. Finally,
the persistence is found to behave like
\begin{align*}
G&\sim\exp[-(t/t_0)^{(1-\al)/(1+\al)}]\times t^{-(3\al-1)/2(\al+1)}
\end{align*}
with $t_0\propto
\nu^{2/(1+\al)}4^{(\al-1)/(\al+1)}/(1-\al)\times[\zeta^{2\al}+\zeta^{\al-1}]$
($\zeta=(1-\al)/2\al$). The coefficient of proportionality for $t_0$ must
be of order $1$.
\begin{figure}
  \onefigure[scale=0.66]{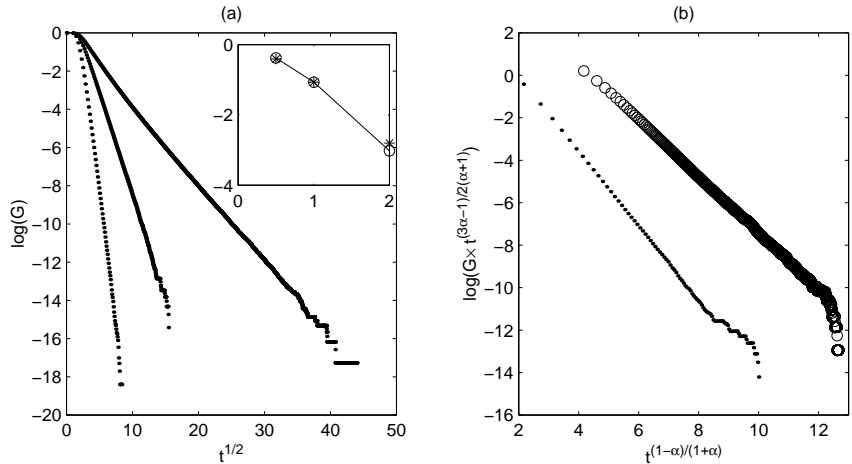}
  \caption{Persistence distribution numerically computed for 
    $A(x)=-\nu x^{-\al}$ with $0<\al<1$. (a) $\al=1/3$ and
    $\nu=0.5,1,2$. The inset shows the $\nu$-dependence of the coefficient
    $t_0$ (stars) compared to  the prediction $\nu^{2/(1+\al)}$ of the
    theory (circles). (b)
    $\al=0.5$ (dots) and $\al=0.75$ (circles) (in that panel, the
values of $t_0$ are of order $0.12$). The ultimate diverging tails
    of these curves are artefacts due to a poor statistics in these regions.}
  \label{figverif}
\end{figure}

To check the validity of our theory, we have numerically computed the
persistence distribution for the last case ($\al<1$) :  figure
\ref{figverif} (a) shows three different curves of $\log(G)$ versus $\sqrt{t}$
for  $\al=1/3$ (note that in that case, the persistence is purely a
stretched exponential, with exponent $1/2$) and
different values of $\nu$ ; the inset shows the $\nu$ dependence of the slopes of
these curves, which  is well predicted by the expression of
$t_0$ given by the above formula. Panel (b) of the figure shows two others examples of
$\log(t^{(3\al-1)/2(\al+1)}G)$ plotted as function of $C^{st}\times
t^{(1-\al)/(1+\al)}$, emphasizing again that  the linear  behaviour
analytically derived is fully satisfied.

\bigskip

As a last example, let us consider now the case of an \textit{underdamped}
particle in a potential 
well, submitted to a damping $\ga(E)$, \textit{a priori} function of its energy. Kramers \cite{kramers,htb} had shown that the fast angle
variable can be eliminated, leading to an effective Fokker-Planck equation
for its energy $E$, with
$A=\frac{\omega(E)}{2\pi}[k_BT\pa_E(\ga(E)I(E))-\ga(E)I(E)]$ and
$B=2k_BT\frac{\omega(E)}{2\pi}\gamma(E)I(E)$ (where $I(E),\omega(E)$ are the action
and the pulsation of the trajectory). If the damping vanishes rapidly enough as $E$ increases, the problem  belongs to the ``gapless''
category. For instance, if one considers the case of an harmonic potential
($\omega=\omega_0,I(E)=2\pi E/\omega_0$), with a damping $\gamma(E)=E^{-\al}$
(for $E$ sufficiently high), one has that, if $\al>1$, the Schr\"odinger
potential goes to zero at infinity, and the persistence is asymptotically
$G\sim t^{3/2\al-1}\exp-(t/t_0)^{\al^{-1}}$. This class of situations is
physically particularly relevant to the cases of non linear coupled oscillators: the frequency shift between two adjacent oscillators having
different energies leads to a enormous slowing down of the diffusion of the
energy \cite{these,tsiau}, described in a mean-field model by such a vanishing
damping.

\section{Correlation functions}

It is interesting to remark that this method can be
applied to the calculation of tails of correlation functions, for systems
belonging to the appropriate ``gapless'' class. The hypothesis that a
correlation function exists implies that a real equilibrium is reachable by
the system ; with this restriction,  the correlation
function of the variables is \cite{gardiner}
\begin{align}
  <x(t)x(0)>&=\int_0^\infty d\la\ e^{-\la t}\left(\int dx\  x\sqrt{p_{st}(x)}\psi_\la(x)\right)^2.
\end{align}
It is clear that an analogous derivation can be performed on that formula,
in order to extract its asymptotic behaviour
with, nevertheless,
a slight difference in the definition of the $\psi_\la$ : the boundary
condition $\psi_{\la}(x_0)=0$ does no longer exist, and the $\psi_0$ function is
now a real eigenfunction equal to $\sqrt{p_{st}}$.

\section{Conclusion}

In this letter, we have presented a general method to derive analytically the
asymptotic behaviour of the persistence probability, for a large number of
markovian processes in one dimension : we have shown that it is possible to classify
markovian processes in three categories ; the first is characterised by an
exponential extinction of the persistence, a second one, quite marginal, whose
treatment is beyond the scope of this letter, and the third, which we
called ``gapless'', in reference to the structure of the spectrum of an
associated Schr\"odinger operator. For this third class, we have introduced
a systematic procedure to obtain the persistence distribution at large
times, and we have tested the validity of the procedure by two
complementary ways : we compared the results of our method to  exactly
solvable models, and  the theoretical predictions to  numerical
simulations,  when the exact result is not known. Both comparisons have
shown an impressive agreement.

It is interesting to note that this method could be presumably  extended to
the treatment of multidimensional cases, as soon as the corresponding
Fokker-Planck equation can be mapped with an appropriate change of
variables on a Schr\"odinger equation. In this extended version of the theory,  relevant
quantities like $x_\la$ would become functions of the solid angle of the
parameter space, leading
probably to a more complicated behaviour of the persistence. This extension
could therefore  be an original way to study non markovian cases, since
 memory effects can always be interpreted as an elimination of
additional variables. 

\acknowledgments

I thank T.Dauxois and A.Alastuey for comments and discussions.

\end{document}